# Electrical Behavior of Downburst-Producing Convective Storms over the High Plains

Kenneth L. Pryor
Center for Satellite Applications and Research (NOAA/NESDIS)
Camp Springs, MD

**Abstract**

A great body of research literature pertaining to microburst generation in convective storms has focused on thermodynamic factors of the pre-convective environment as well as storm morphology as observed by radar imagery. Derived products based on Geostationary Operational Environmental Satellite (GOES) sounder data have been found to be especially useful in the study of thermodynamic environments. However, addressed much less frequently is the relationship between convective storm electrification, lightning phenomenology and downburst occurrence. Previous research in lightning production by convective storms has identified that electrification, phenomenology (i.e. flash rate, density), and polarity are dependent upon the thermodynamic structure of the ambient atmosphere, especially vertical moisture stratification. Thus, relevant parameters to describe the thermodynamic setting would include convective available potential energy (CAPE), due to its influence on updraft strength, and moisture characteristics of the boundary layer, due to its relationship to precipitation physical processes. It has already been addressed that buoyant energy and moisture stratification are important factors in convective storm development and downburst generation. This research effort investigates and derives a qualitative relationship between lightning phenomenology in convective storms and downburst generation. Downburst-producing convective storms will be analyzed by comparing pre-convective environments, as portrayed by GOES microburst products, storm morphology, as portrayed by radar imagery, and electrical behavior, as indicated by National Lightning Detection Network (NLDN) data. In addition, this paper will provide an update to validation efforts for the GOES Wet Microburst Severity Index (WMSI) and Hybrid Microburst Index (HMI) products to include data from both the 2005 and 2006 convective seasons. The case study presented will demonstrate the effectiveness of the coordinated use of the GOES WMSI and HMI products during convective storm events over the southern High Plains.

## 1. Introduction

A great body of research literature pertaining to microburst generation in convective storms has focused on thermodynamic factors of the pre-convective environment as well as storm morphology as observed by radar imagery. Derived products based on Geostationary Operational Environmental Satellite (GOES) sounder data have been found to be especially useful in the study of thermodynamic environments. However, addressed much less frequently is the relationship between convective storm electrification, lightning phenomenology and downburst generation. Previous research in lightning production by convective storms has identified that electrification, phenomenology (i.e. flash rate, density), and polarity are dependent upon the thermodynamic structure of the ambient atmosphere, especially vertical moisture stratification. Thus, relevant parameters to describe the thermodynamic setting would include convective available potential energy (CAPE), due to its influence on updraft strength, and moisture characteristics of the boundary layer, due to its relationship to precipitation physical processes. It has already been addressed that buoyant energy and moisture stratification are important factors in convective storm development and downburst generation.

The basic concept of charge generation and separation in convective clouds maintains that the presence of strong updrafts and the resulting development of precipitation are instrumental in the formation of an electric field of sufficient intensity for lightning discharge. In a convective cloud that builds to a height well above the freezing level, intense vertical motion will also loft large amounts of ice crystals to near the storm top as well as into the anvil, if present. At the same time, precipitation, in the form of graupel or aggregates, will begin to descend within the convective cloud through the loading process. The interaction between ice crystals and larger precipitation particles (i.e. graupel, aggregates) will result in the acquisition of opposite electric charge between the lighter ice particles and the heavier precipitation particles, establishing a dipole and an electric field intensity necessary for electrical breakdown and subsequent lightning initiation (Saunders 1993). The descending precipitation within the storm is also vital for the development of convective downdrafts, accelerated as drier air from outside the convective storm cell is entrained. Thus, the physical process responsible for the initiation of lightning within a convective storm is also believed to be instrumental in the initiation of convective downdrafts that eventually produce downbursts at the surface. Pryor and Ellrod (2005) developed a Geostationary Operational Environmental Satellite (GOES) sounder-derived wet microburst severity index (WMSI) product to assess the potential magnitude of convective downbursts, incorporating convective available potential energy (CAPE) as well as the vertical theta-e difference (TeD) between the surface and mid-troposphere. In addition, Pryor (2006) developed a GOES Hybrid Microburst Index (HMI) product intended to supplement the use of the GOES WMSI product over the United States Great Plains region. The HMI product infers the presence of a convective boundary layer (CBL) by incorporating the sub-cloud temperature lapse rate as well as the dew point depression difference between the typical level of a convective cloud base and the sub-cloud layer. Thus, the WMSI algorithm is designed to parameterize the physical processes of updraft and downdraft generation within the convective storm cell, while the HMI algorithm describes the moisture stratification of the sub-cloud layer that may result in further downdraft acceleration, eventually producing a downburst when the convective downdraft impinges on the earth's surface.

Accordingly, lightning data imagery, generated by a program that ingests the data from the [National Lightning Detection Network (NLDN)](#) to be plotted and mapped by Man computer Interactive Data Access System (McIDAS), has been collected for several severe convective storm events that occurred in Oklahoma during the summer of 2006. For each downburst occurrence, NLDN data has been compared to radar and satellite imagery and surface observations of downburst winds. It has been observed that suppressed cloud-to-ground (CG) lightning discharge was associated with downburst occurrence in most cases. Common between the environments of these convective storms was the presence of a relatively deep and dry CBL, as indicated by GOES HMI imagery and surface observations. The reader is referred to MacGorman and Rust (1998) and Uman (2001) for a discussion of convective storm electrification and lightning phenomenology.

It has been suggested that decreased CG flash rates in convective storms may be a consequence of an elevated charge dipole resulting from especially vigorous updrafts. Equation 3.1 in Rakov and Uman (2003), derived from Coulomb's law, describes the quantitative relationship between electric field intensity, $\mathbf{E}$, the height (H) of the lower (typically negative) charge center above the surface and the charge magnitude (Q) of the lower charge center: $\mathbf{E} \sim Q/H^2$. As dictated by Rakov and Uman (2003), the upward displacement of the lower charge center typically results in a reduced electric field intensity between the convective cloud and the surface. Another effect of intense convective updrafts would be to increase precipitation content within the convective cell, thereby enhancing the effect of precipitation loading (Doswell 2001; Wakimoto 2001). Once the process of precipitation loading has initiated a downdraft, entrainment of dry (or low theta-e) air in the mid levels of the storm cell would enhance downdraft strength by the process of evaporative cooling. Thus, periods of enhanced updraft intensity within a convective storm could be associated with an increased likelihood for downburst generation, and, hence, imply a relationship between CG flash rates and downburst occurrence.

This research effort investigates and derives a qualitative relationship between lightning phenomenology in convective storms and downburst occurrence. Downburst-producing convective storms will be analyzed by comparing pre-convective environments, as portrayed by GOES microburst products, storm morphology, as portrayed by radar imagery, and electrical behavior, as indicated by NLDN data. In addition, this paper will provide an update to validation efforts for the GOES Wet Microburst Severity Index (WMSI) and Hybrid Microburst Index (HMI) products from 2006 convective season. The case study presented will demonstrate the effectiveness of the coordinated use of the GOES WMSI and HMI products during convective storm events over the southern High Plains.

## 2. Methodology

Data from the GOES HMI and WMSI products was collected over Oklahoma from 1 June to 31 August 2006 and validated against conventional surface data. GOES HMI and WMSI images were generated by Man computer Interactive Data Access System (McIDAS) and then archived on an FTP server (ftp://ftp.orbit.nesdis.noaa.gov/pub/smcd/opdb/wmsihmiok/). Validation from the 2005 convective season identified a statistically significant correlation over the Oklahoma Panhandle region between GOES-derived WMSI values and observed surface convective wind gust magnitude when the WMSI product was used in conjunction with the GOES HMI product. Therefore, within the State of Oklahoma, the panhandle region was selected as an area of study due to the wealth of surface observation data provided by the [Oklahoma Mesonet](#) (Brock et al. 1995), a thermodynamic environment typical of the High Plains region during the warm season, proximity to the dryline, and relatively homogeneous topography. The High Plains region encompasses the Oklahoma Panhandle that extends from 100 to 103 degrees West (W) longitude. The ground elevation on the panhandle increases from near 2000 feet at 100W longitude to nearly 5000 feet at 103W longitude (Oklahoma Climatological Survey 1997). Atkins and Wakimoto (1991) discussed the effectiveness of using mesonet observation data in the verification of the occurrence of downbursts. Downburst wind gusts, as recorded by Oklahoma Mesonet observation stations, were measured at a height of 10 meters (33 feet) above ground level. In addition, GOES proximity sounding profiles, most representative of the preconvective environment, were collected for each downburst event. Also, in order to assess the predictive value of GOES microburst products, GOES data used in validation was obtained for retrieval times one to three hours prior to the observed surface wind gusts, assuming that no change in environmental static stability and air mass characteristics between product valid time and time of observed downburst had occurred. Correlations between GOES HMI and WMSI values, and between GOES WMSI values and observed surface wind gust velocities were computed for the validation period. Hypothesis testing was conducted, in the manner described in Pryor and Ellrod (2004), to determine the statistical significance of linear relationships between observed downburst wind gust magnitude and microburst risk values. In order for the downburst observation to be included in the validation data set, it was required that precipitation be observed by each respective mesonet station and the parent convective storm cell of each downburst be located overhead at the time of downburst occurrence. This criteria were based on the definition of wet and hybrid microbursts as stated in Wakimoto (1985) and Atkins and Wakimoto (1991). An additional criterion for inclusion of convective wind gust observations into the validation data set was a measured wind gust of at least F0 intensity (35 knots) on the Fujita scale (Fujita 1971). Wind gusts of 35 knots or greater are considered to be operationally significant for transportation, especially boating and aviation.

Next Generation Radar (NEXRAD) base reflectivity imagery (level II) from National Climatic Data Center (NCDC) was utilized to verify that observed wind gusts were associated with convective storm activity. NEXRAD images were generated by the NCDC Java NEXRAD Viewer (Available online at http://www.ncdc.noaa.gov/oa/radar/jnx/index.html). Another application of the NEXRAD imagery was to infer microscale physical properties of downburst-producing convective storms. Particular radar reflectivity signatures, such as the bow echo and rear-inflow notch (RIN)(Przybylinski 1995), were effective indicators of the occurrence of downbursts. The height of 30 dBZ reflectivity, assumed to represent the upper extent of the mixed phase precipitation core (i.e. graupel, ice crystals, supercooled water), was estimated for each storm of interest by inspecting various radar elevation scans in volume coverage pattern (VCP) 11, 12, or 21. In addition, echo tops (ET) and hail index (HI)(level III) data were collected and analyzed to infer convective storm updraft intensity, based on the premise that higher echo tops and the presence of hail are associated with more intense convective updrafts. At or near the time of each downburst occurrence, the storm echo top and maximum height of 30 dBZ reflectivity was noted and compared to the level of the the -20C isotherm as observed in the respective GOES proximity soundings. Based on a review of numerous studies, Saunders (1993) considered the level of the -20C isotherm to coincide with the upper limit of a transition zone between updrafts and downdrafts in a typical warm-season convective storm, believed to be a region highly favored for the development of a lower electric charge center due to the interaction of precipitation particles in various phases. Electric charging is typically expected between graupel and ice crystals in the presence of supercooled water between the levels of the -10C and -20C isotherms. The resulting region of high graupel concentration comprises the lower charge center while an upper charge center develops as a result of the lofting of ice crystals by convective updrafts to the upper levels of the convective storm cell. He also noted that for electrification to occur in a convective storm, it was necessary for cloud tops to exceed the level of the -20C isotherm.

In addition, National Lightning Detection Network (NLDN) data was collected, mapped by McIDAS and implemented into GOES microburst product imagery. In product imagery, color-coded lightning markers are graduated in ten-minute intervals during the hour following the valid time of each product. As shown in the example in Figure 3, the plotting of cloud-to-ground (CG) lightning progresses in ten-minute periods during each hour, represented by markers colored white, blue, red, green, yellow and cyan, respectively. The number of cloud-to-ground (CG) lightning discharges was estimated by visual inspection for the ten-minute period encompassing downburst observation associated with each convective storm. Validation data set selection criteria were further modified to require that a downburst signature as identified in radar reflectivity imagery (i.e. bow echo apex, RIN) be collocated with the observation of peak convective winds. This location was then compared with the plot of CG lightning markers as displayed in microburst product imagery. This process was necessary to isolate downburst location from surrounding convective storm activity to obtain an accurate CG lightning flash count for each downburst event. Investigation of the elevated dipole hypothesis necessitated that storm echo top and 30 dBZ echo top heights be compared to the height of the -20C isotherm for each downburst event.

3. **Case Study: 13 August 2006 Downbursts**

During the afternoon of 13 August 2006 clusters of convective storms developed along a diffuse frontal boundary that extended from Kansas southwestward through the Oklahoma Panhandle. Two particularly intense downbursts occurred in association with multi-cellular storms that evolved over Cimarron and Texas Counties in the western panhandle. Table 1 presents relevant details pertaining to downburst and CG lightning activity associated with the convective storms. The early afternoon (1900 UTC) GOES HMI product shown in Figure 3 indicated the presence of a very unstable convective boundary layer over the Oklahoma Panhandle. The meteogram from Kenton mesonet station in Figure 1 displayed the evolution of the convective mixed layer through the afternoon with the surface dew point depression reaching a maximum of 37F (20C) by 1900 UTC. Stull (1988) noted that a large surface dewpoint depression (> 17C) is associated with a well-developed CBL. A similar trend in Figure 1 was also noted in the meteogram from Goodwell with a slightly larger surface dewpoint depression (22C) prior to downburst occurrence. However, a major difference in afternoon convective boundary layer evolution between Kenton and Goodwell portrayed in the meteograms was a frontal passage at Goodwell shortly before 1900 UTC marked by a significant temperature decrease, dewpoint increase and wind shift. It is evident that intense solar heating after the frontal passage resulted in strong thermal circulation and mixing in the boundary layer. Inspection of GOES sounding profiles in Figure 2 from Clayton, New Mexico, near Kenton, and Guymon, Oklahoma, near Goodwell, revealed "inverted-v" profiles with significant CAPE and an elevated mixed layer depth typical of the warm-season environment over the High Plains. The surface analysis at 1900 UTC (not shown) located the cold front over the central Oklahoma Panhandle by a wind shift and dewpoint discontinuity. Microburst product imagery in Figure 3 at 1900 UTC supported the location of the front identified in the surface analysis as a line of developing convective storm activity.

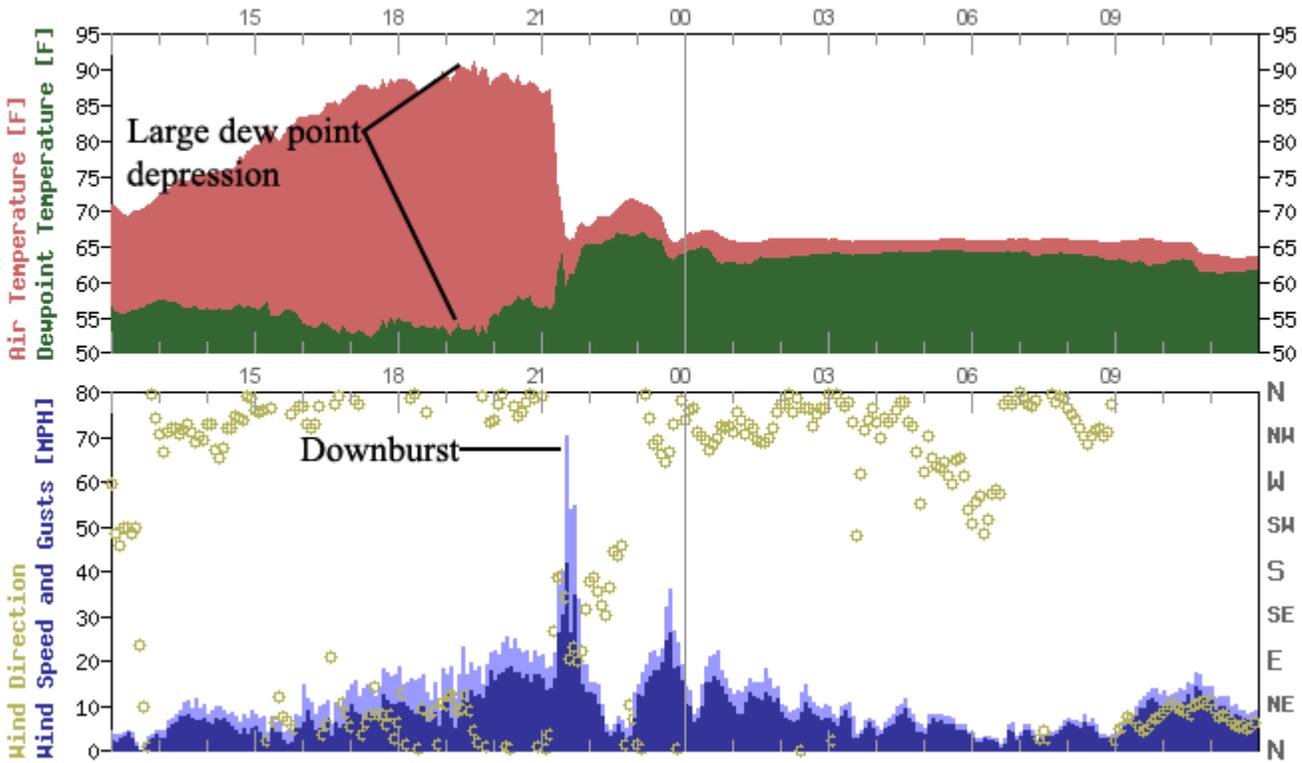
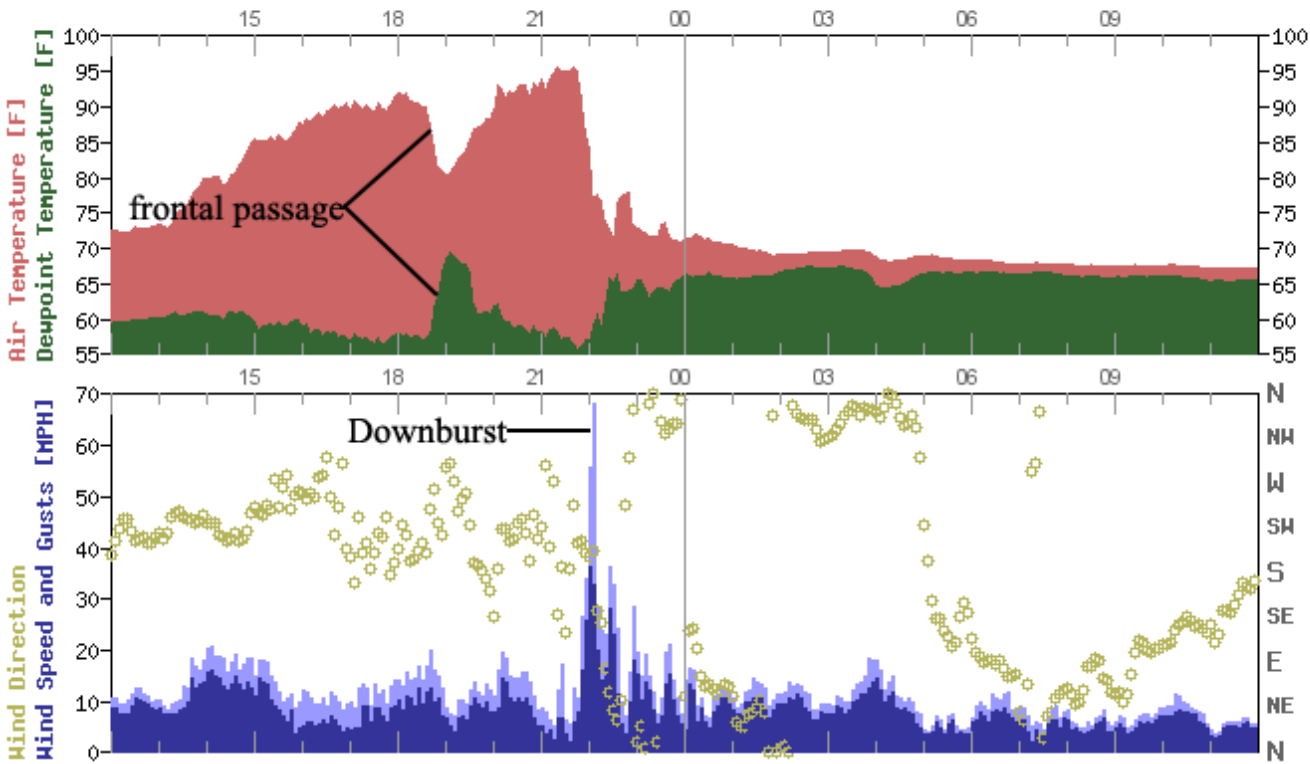

Figure 1. Oklahoma Mesonet meteograms at Kenton(top) and Goodwell (bottom).

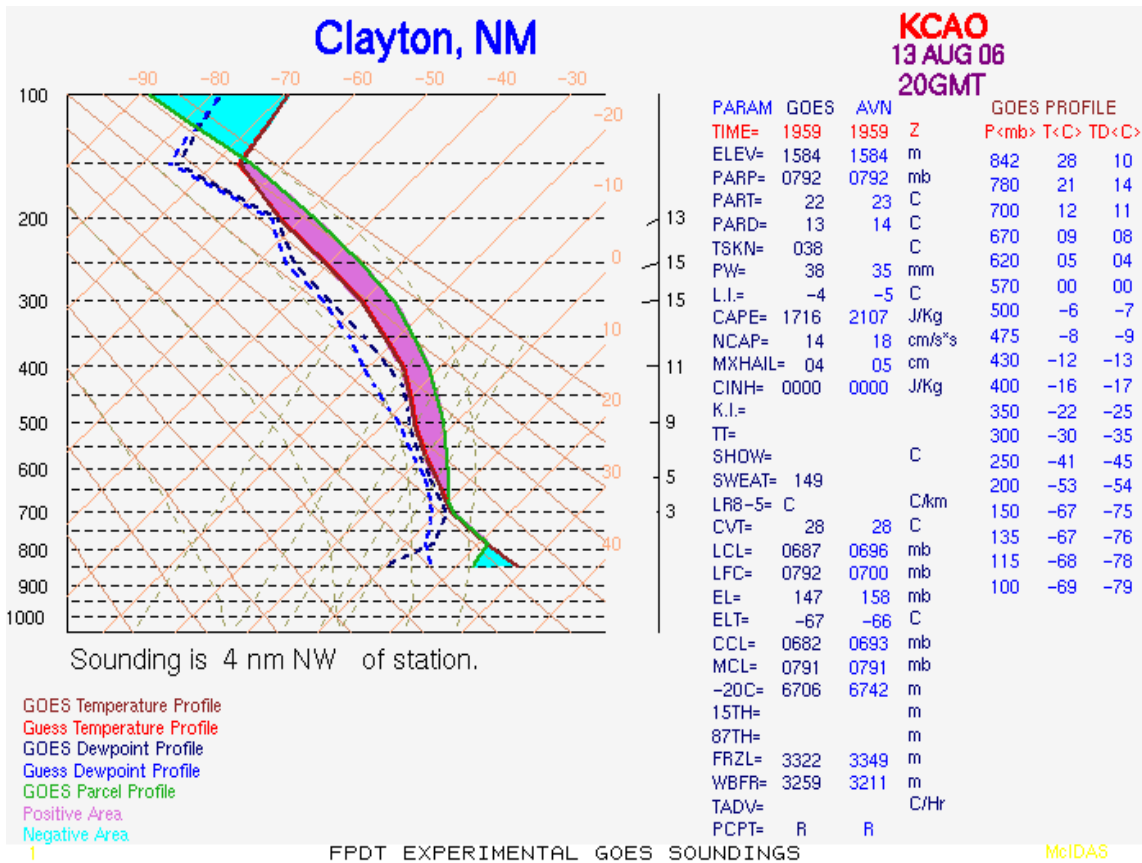

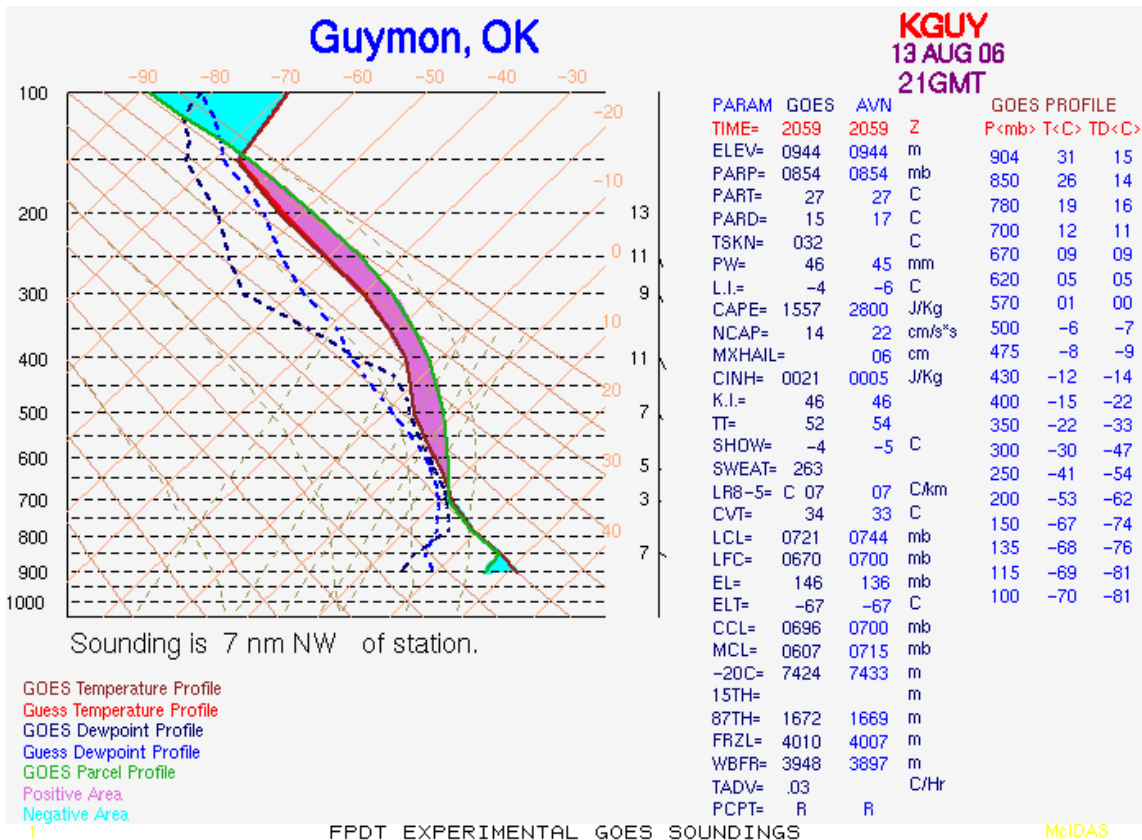

Figure 2. GOES soundings at Clayton, New Mexico at 2000 UTC (top) and Guymon, Oklahoma at 2100 UTC (bottom) 13 August 2006.

A multi-cellular convective storm propagated northeastward over western Cimarron County between 2100 and 2200 UTC. The first severe downburst observed over the Oklahoma Panhandle occurred as an embedded cell evolved into a bow echo configuration. Figure 5, radar reflectivity imagery, displayed a well-defined bow echo and RIN over Kenton at the time of downburst occurrence. Also noteworthy is the close correspondence between the location of maximum echo tops and the location of the bow echo apex associated with the downburst. Further analysis of level II and level III NEXRAD data revealed similar morphology characteristics between the bow echo and attendant downburst that was observed at Goodwell (Figure 6). This observation suggests that the presence of strong updrafts was a possible contributor to the development of an elevated electric charge dipole within the downburst-producing convective storms. Meteograms provided further evidence of the occurrence of downbursts. Specifically, well-defined peaks in wind speed as well as significant temperature decreases (Atkins and Wakimoto 1991) were effective indicators of downburst occurrence, as displayed by Kenton and Goodwell meteograms. The close spatial correspondence between peak wind observation and bow echo signature location suggest that downburst winds observed at Kenton and Goodwell were in close proximity to the point of impact at the surface.

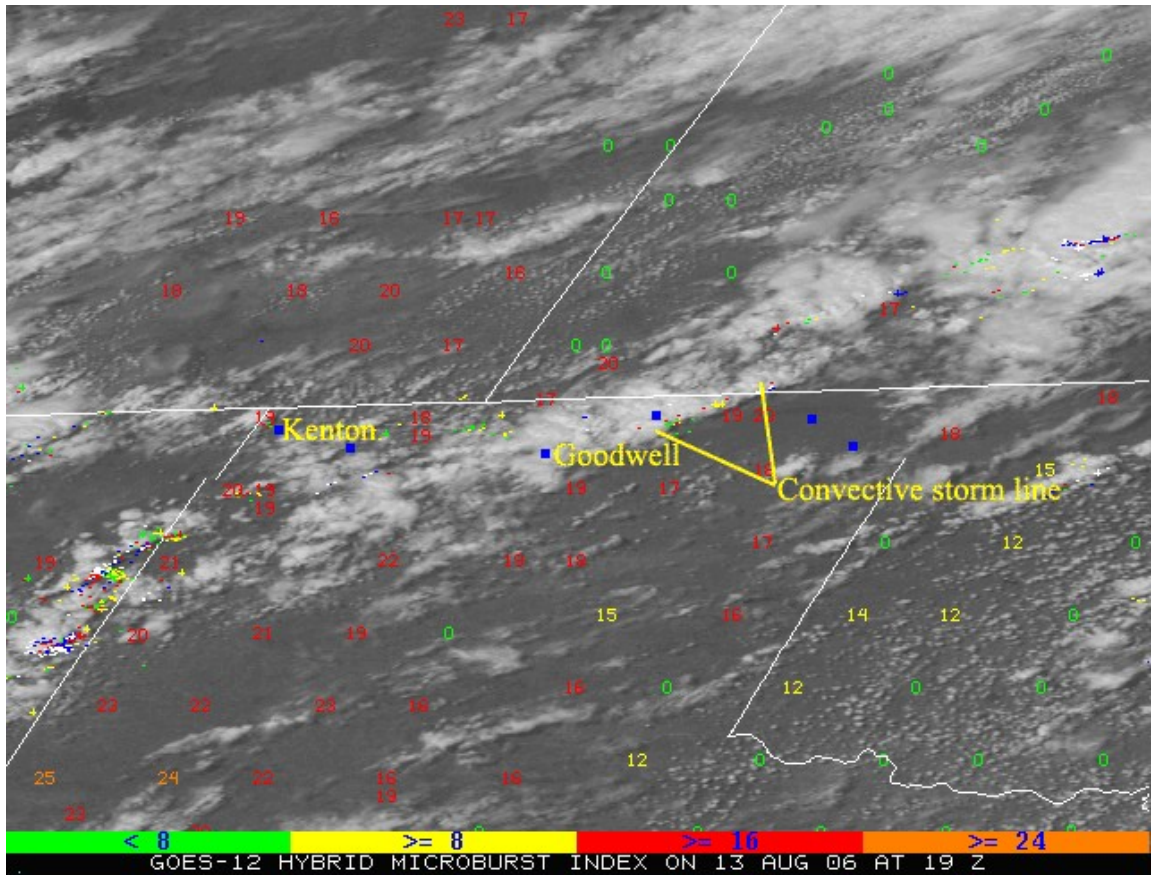

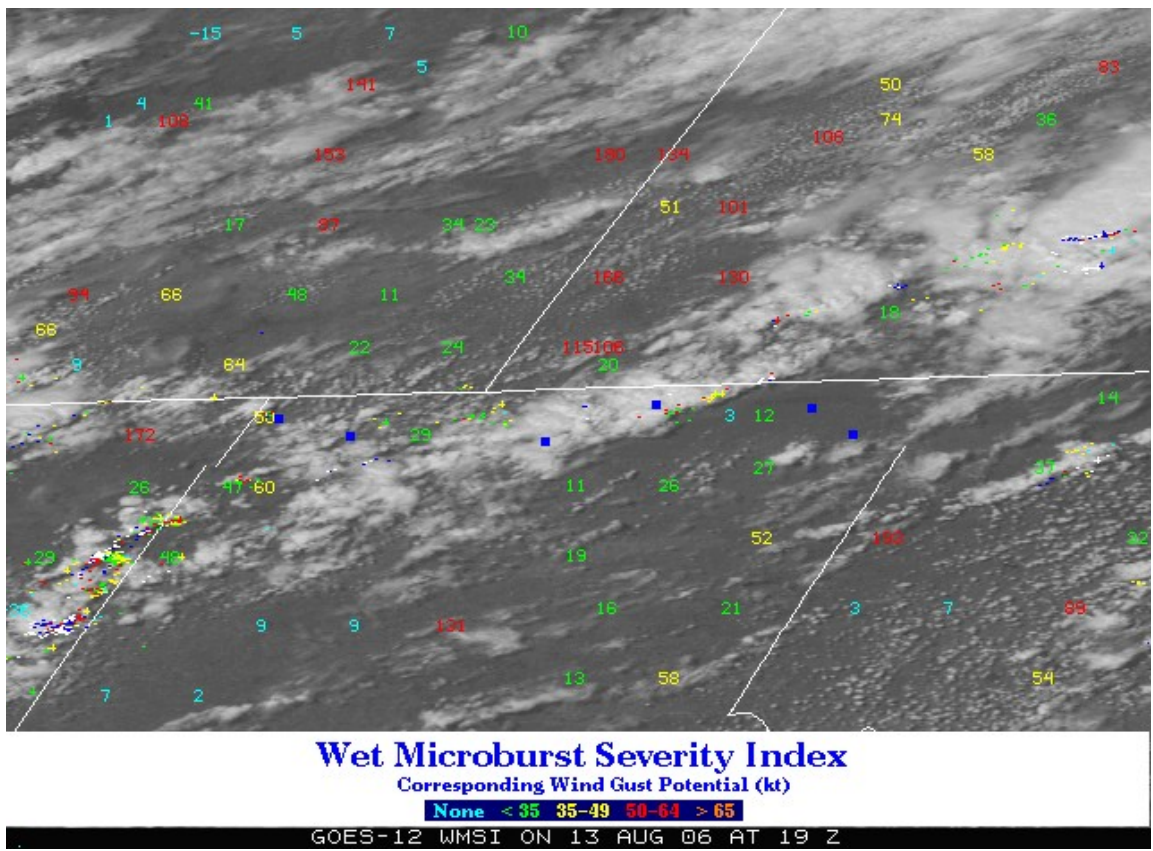

Figure 3. GOES HMI (top) and GOES WMSI (bottom) at 1900 UTC 13 August 2006.

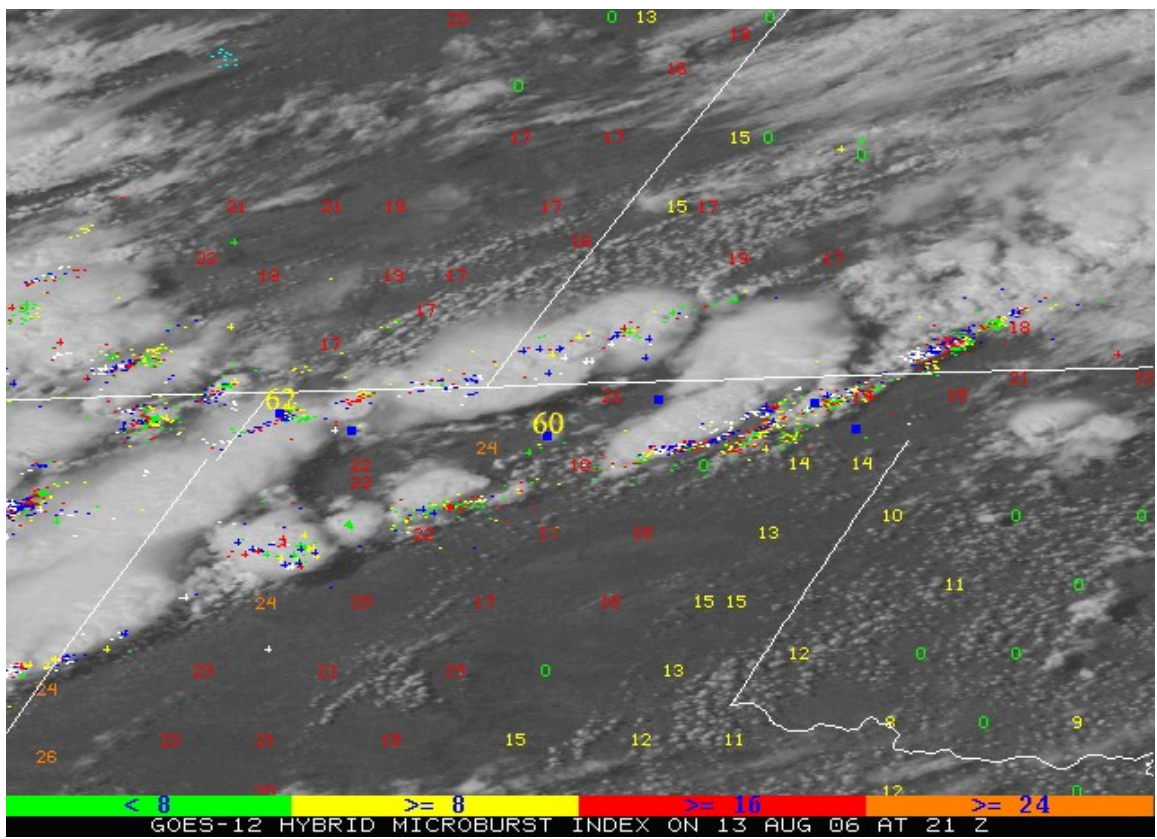

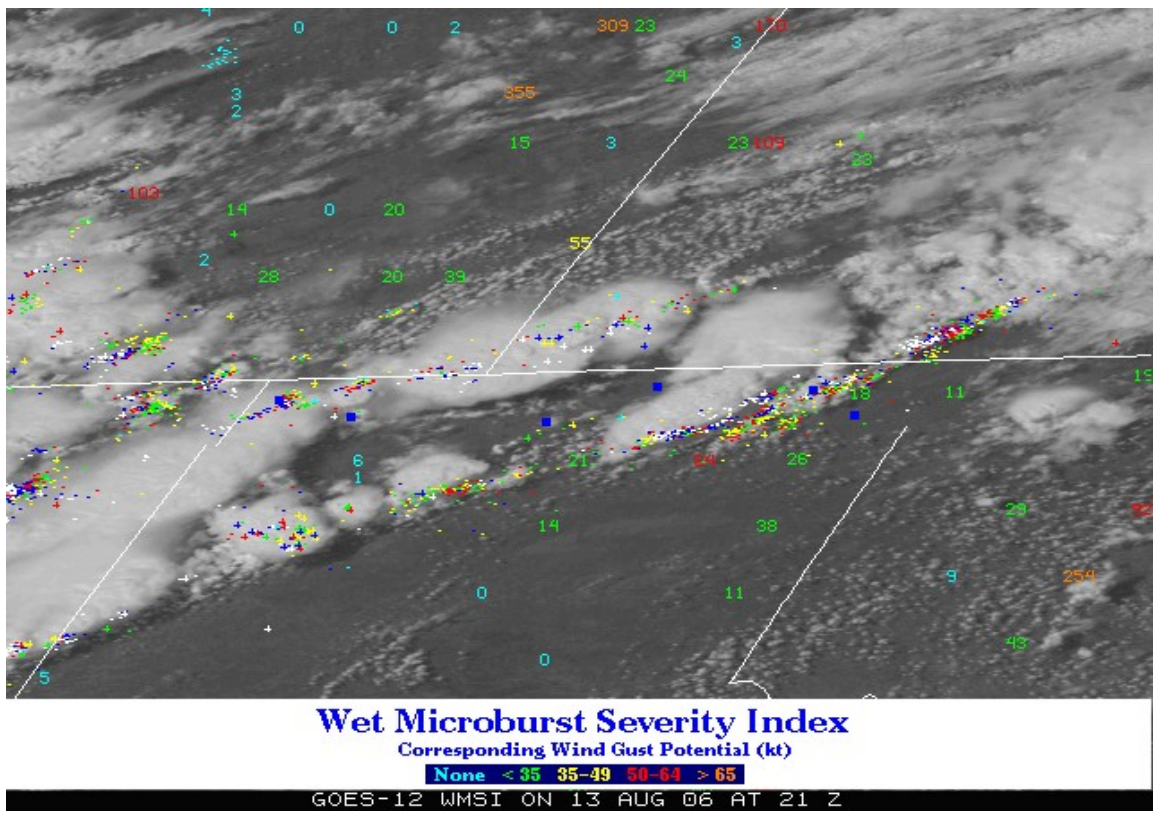

Figure 4. GOES HMI (top) and GOES WMSI (bottom) at 2100 UTC 13 August 2006. Downburst wind gust magnitude observed at Kenton and Goodwell mesonet stations, respectively, are indicated in HMI image.

The convective boundary layer was maintained over the panhandle through the afternoon as indicated in Figure 4 by the 2100 UTC HMI and WMSI products. The HMI image displays elevated risk values (> 18) over the central Oklahoma Panhandle in the vicinity of the frontal boundary. Also apparent in the WMSI image is the presence of static instability as represented by elevated risk values (>20).

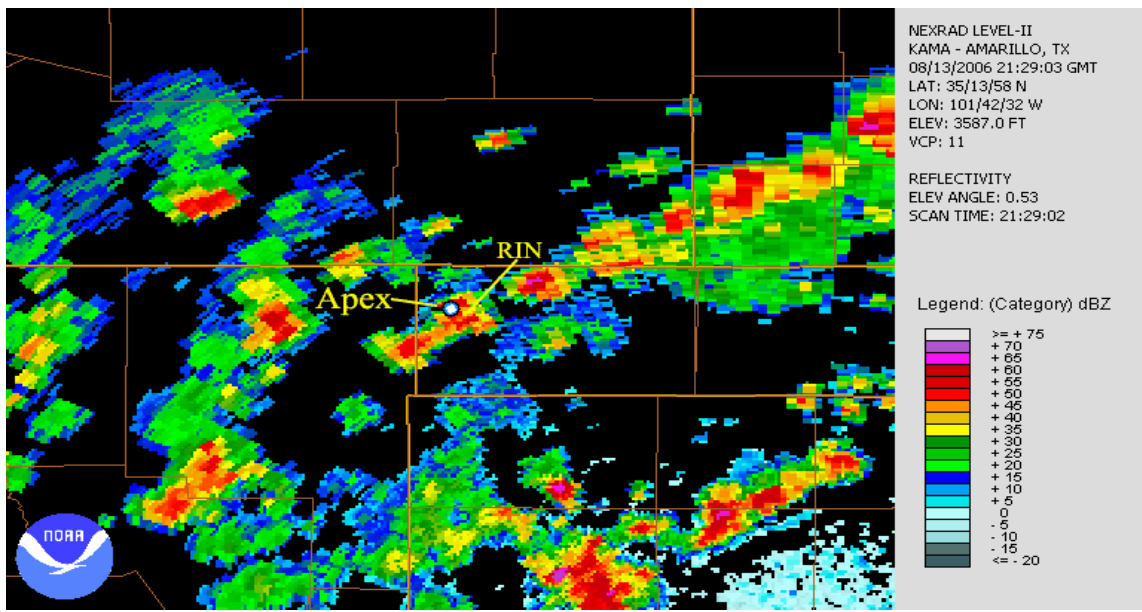

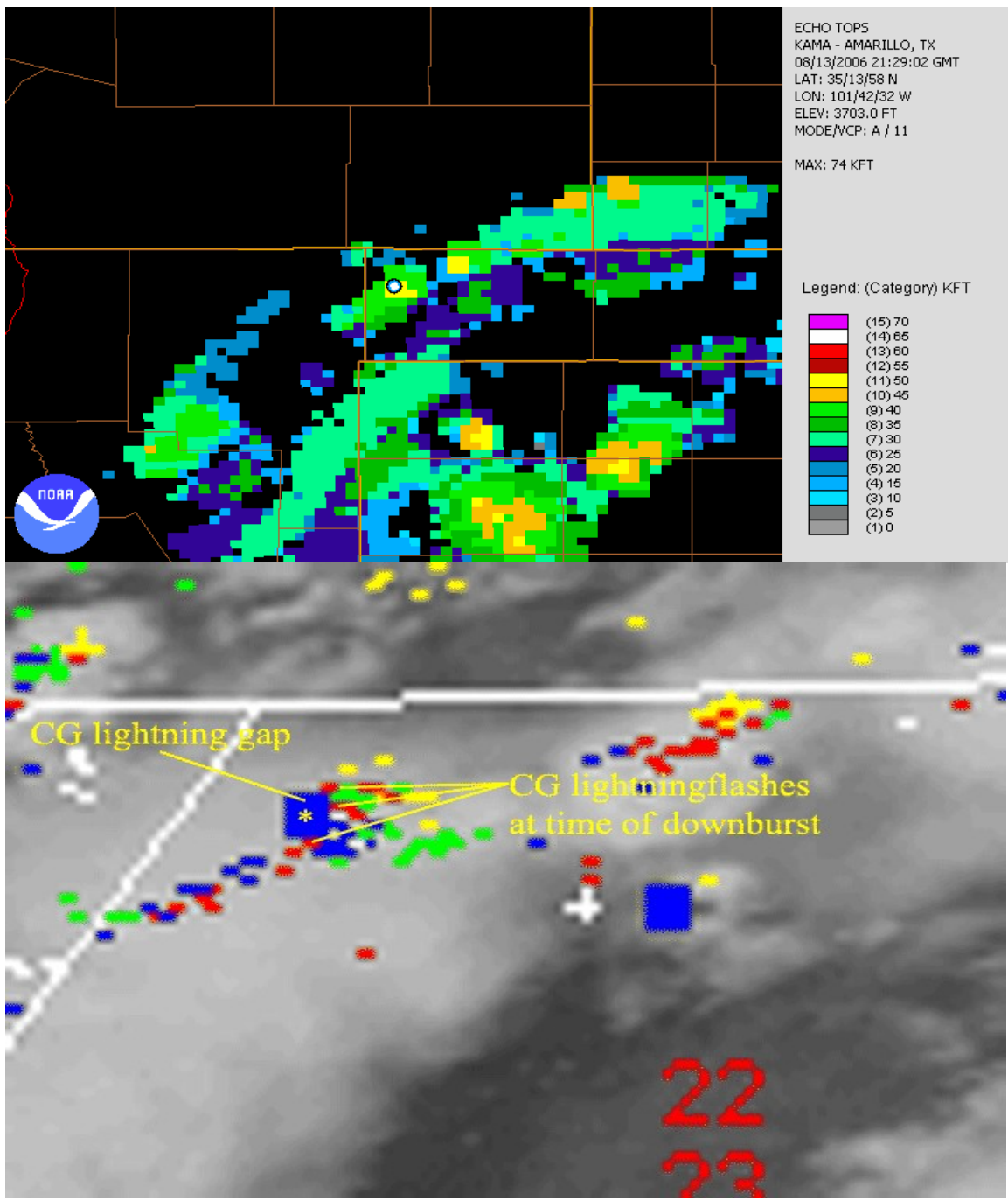

Figure 5. NEXRAD reflectivity (top) and echo tops (center) at 2129 UTC 13 August 2006; magnified HMI image at 2100 UTC 13 August 2006. Location of Kenton mesonet station in radar images is indicated by a blue and white marker and in the HMI image by a yellow asterisk within a blue square.

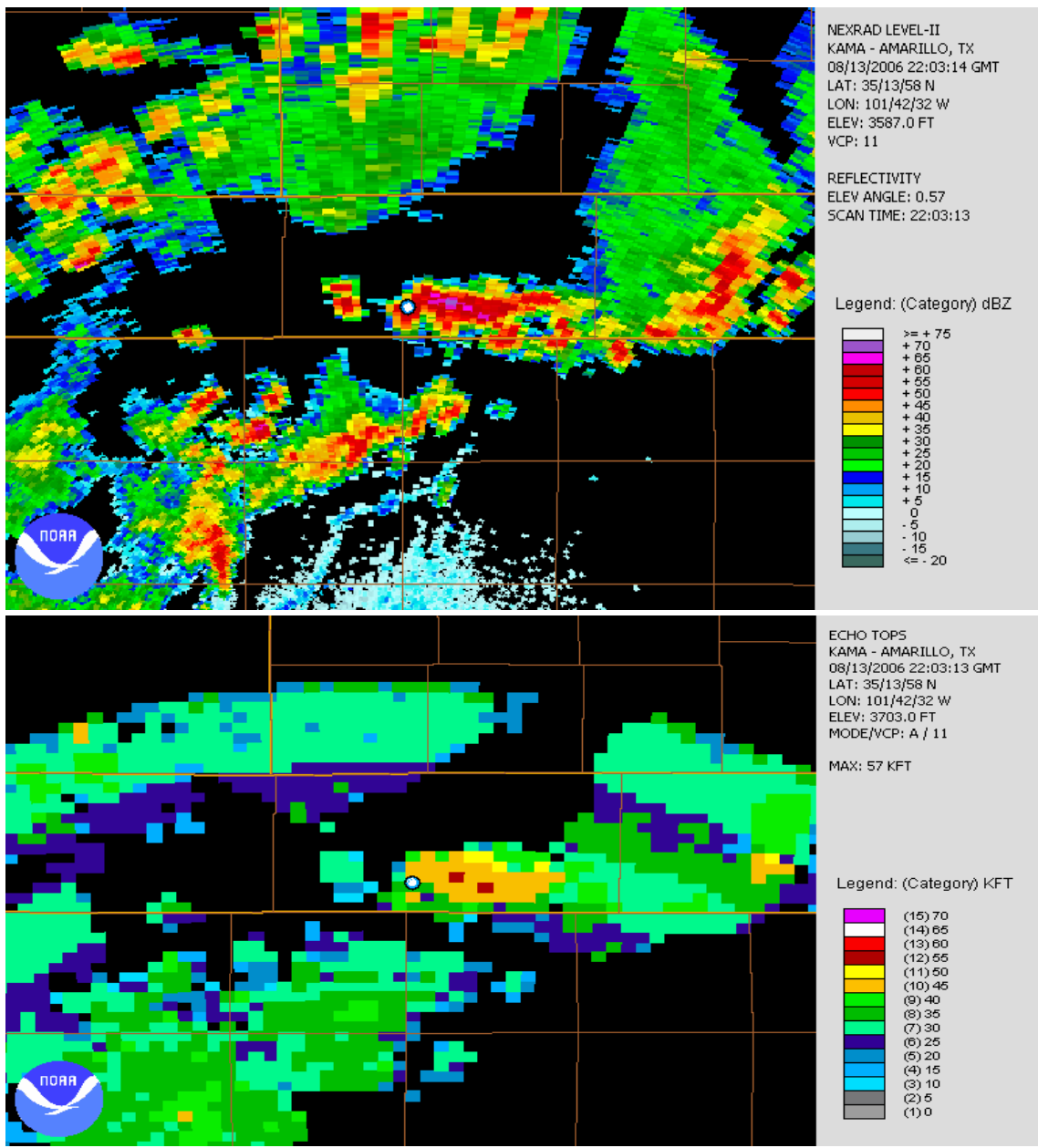

Figure 6. NEXRAD reflectivity (top) and echo tops (bottom) at 2203 UTC 13 August 2006. Location of Goodwell mesonet station is indicated by a blue and white marker.

In addition, cloud-to-ground (CG) lightning strikes that occurred during the one-hour period following the valid time of each product are shown in Figures 3, 4, and 7 in association with the convective storms. Apparent in a magnified microburst product image at 2100 UTC is suppressed CG lightning activity in the vicinity of observation of peak downburst winds at Kenton. This feature, referred to as a CG lightning "gap" in this paper, was also noted in the HMI image at 2200 UTC in association with peak downburst winds observed at Goodwell. Analysis of echo tops as well as elevation scans in VCP 11 from Amarillo, Texas NEXRAD revealed that at the time of downburst observation at Kenton and Goodwell, storm echo tops approached or exceeded 15240 meters (50000 feet), well above the level of the -20C isotherm between 6706 meters (22000 feet) and 7424 meters (24356 feet), as indicated by GOES sounding profiles at Clayton and Guymon, respectively. In addition, 30 dBZ echo top heights significantly exceeded the height of the -20C isotherm, suggesting that the observed CG lightning gaps were most likely the result of an elevated charge dipole.

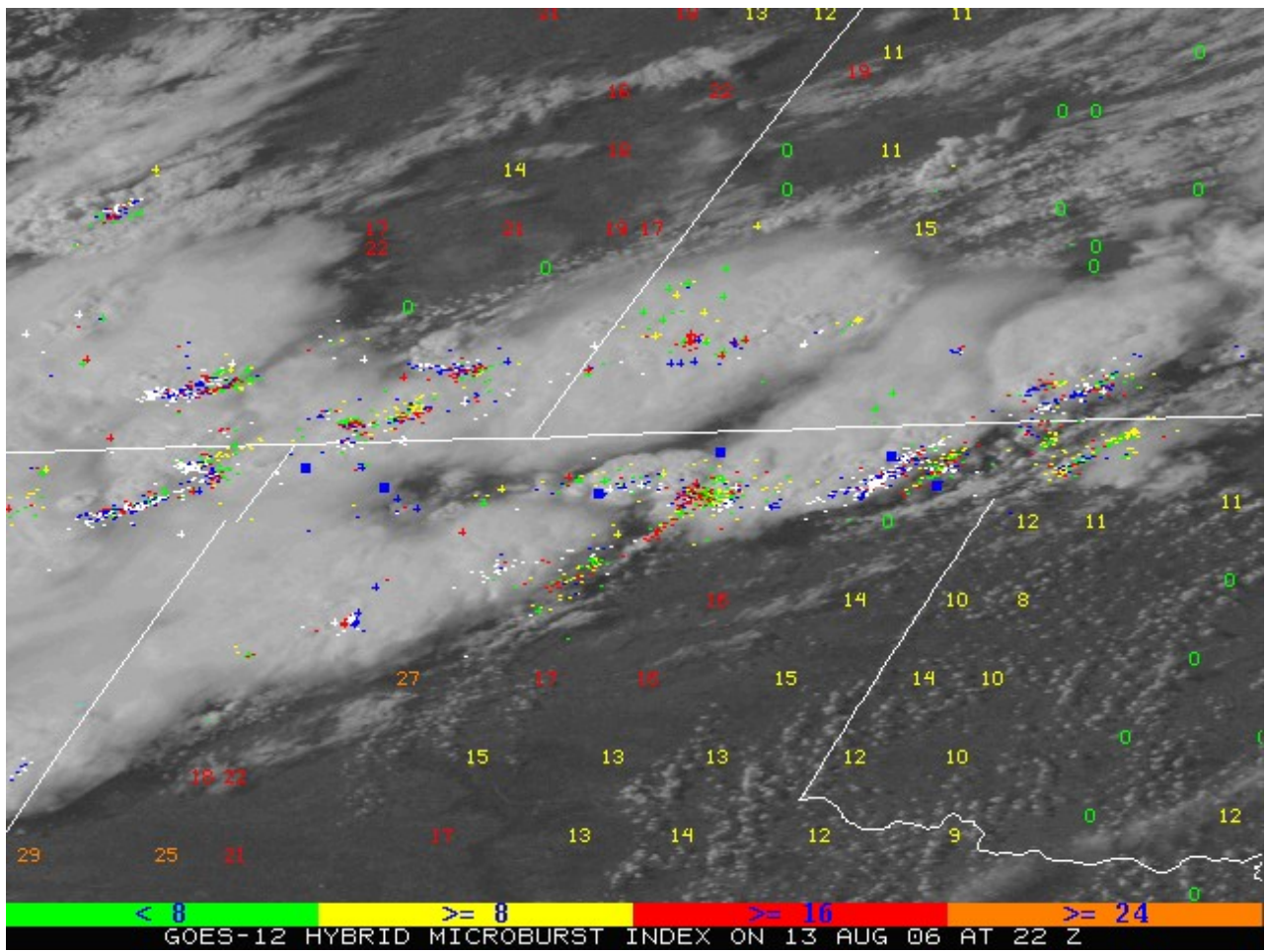

Figure 7. GOES HMI at 2200 UTC 13 August 2006.

| Table 1. 13 August 2006 Downbursts | | | | | | |
|---|---|---|---|---|---|---|
| Time (UTC) | Location | Measured Surface Wind (knots) | HMI | WMSI | CG Lightning Marker Color (at time of downburst) | Image Valid Time (UTC) |
| 2130 | Kenton | 62 | 20 | 50 | Red | 2100 |
| 2205 | Goodwell | 60 | 18 | 21 | White | 2200 |

## 4. Discussion and Conclusions

| Table 2. Analysis of Statistics | | |
|---|---|---|
| Convective Seasons 2005-2006 (N=22) | | |
| | Correlation (r) | t value |
| WMSI to measured wind | 0.55 | 5.36 |
| WMSI to HMI | -0.5 | 5.16 |
| Critical Value | 1.77 | |

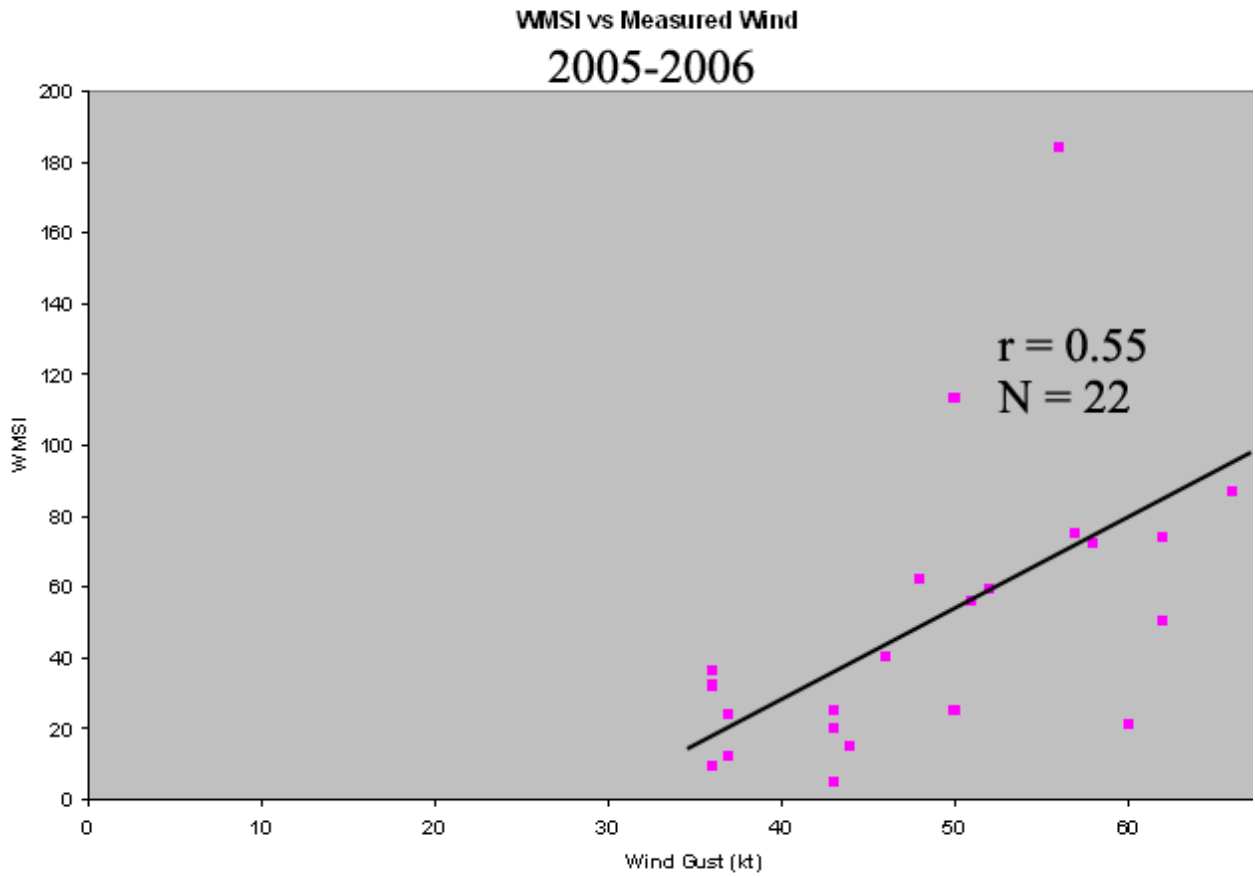

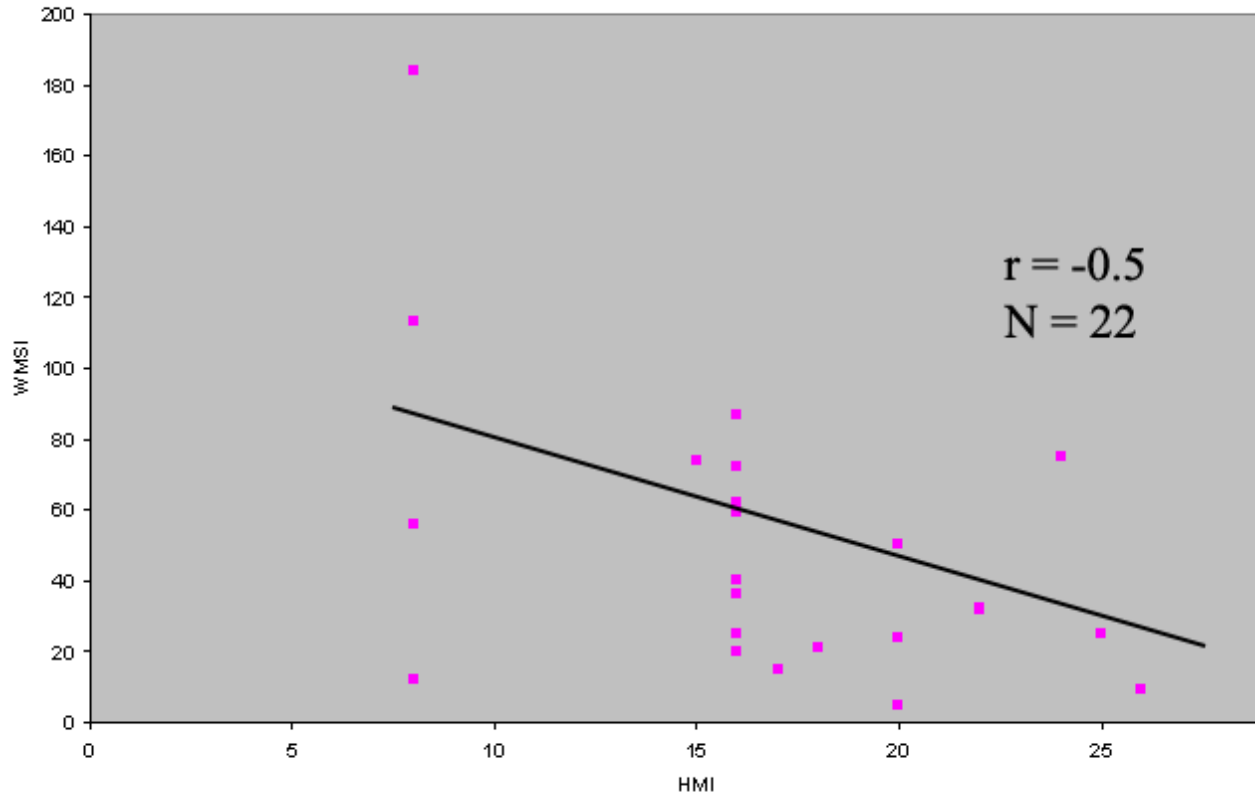

Figure 8. Scatterplots comparing WMSI values to measured peak downburst wind gusts (top) and WMSI to HMI values for each downburst occurrence (bottom).

Analysis of correlation statistics for downburst events over the Oklahoma Panhandle during the 2005 and 2006 convective seasons demonstrates interesting results as portrayed in Figure 8. Hypothesis testing, as conducted in the manner described in Pryor and Ellrod (2004), revealed a statistically significant correlation between WMSI values and measured downburst wind gusts. Also interesting to note is the statistically significant negative correlation between WMSI and HMI values. This inverse proportionality between WMSI and HMI values for convective wind gusts of comparable magnitude likely reflects a continuum of favorable environments for downbursts, ranging from wet, represented by high WMSI values and correspondingly low HMI values, to an intermediate or hybrid regime, characterized by a combination of lower WMSI values and elevated HMI values. The inverse relationship between WMSI and HMI values associated with observed downbursts underscores the relative importance of thermodynamic structure and moisture stratification of the boundary layer in the acceleration of convective downdrafts. The increasing influence of thermal and moisture stratification in the convective boundary layer on downburst wind gust magnitude becomes apparent with higher HMI values. Accordingly, the convective wind gust prediction technique utilizing a combination of WMSI and HMI values proposed in Pryor (2006) was found to be effective when incorporating validation statistical data from the 2006 convective season:

| Table 3. Convective Wind Gust Prediction Matrix | | |
|---|---|---|
| HMI | WMSI | Wind Gusts (kt) |
| <8 | 10 - 49 | < 35 |
|  | 50 - 79 | 35-49 |
|  | > 80 | >50 |
| > or =8 | 10 - 30 | < 35 |
|  | 30 - 49 | 35-49 |
|  | 50 - 79 | >50 |
|  | > 80 | > 50 |
| > or =16 | 10 - 30 | 35 - 49 |
|  | 30 - 49 | >50 |
|  | 50 - 79 | >65 |
|  | > 80 | >65 |

Also of importance, as identified in the statistical analysis, is the relationship between downburst CG lightning flash counts and corresponding WMSI and HMI values. As exemplified in the case study, suppression of CG lightning was observed with a combination of significant WMSI and HMI values, suggesting that moisture characteristics of the boundary layer as well as convective precipitation processes effected storm electrification and the lightning "gaps" associated with observed downbursts. In addition, the relationship between the storm and 30 dBZ echo top heights and downburst CG lightning flash counts was considered. In most cases, the suppression of CG lightning activity was most pronounced where the maximum vertical extent of 30 dBZ reflectivity was observed, most likely coupled to an elevated charge dipole realized by strong convective updrafts. It follows that especially strong updrafts can lift the mixed-phase precipitation core to a height at which the electric field intensity between the cloud and the surface decreases below a critical value for CG lightning discharge. It is also imperative to note that Fujita's model of downburst evolution (Fujita 1985) suggests that the surface location of peak downburst winds is typically displaced from the convective storm downdraft core by up to several kilometers. As the intense downdraft impinges on the surface, the resulting divergent outflow induces a burst or surge of strong and potentially damaging winds emanating from the point of impact. The location of peak surface winds associated with the downburst is expected to be observed near the forward flank of a migratory convective storm and underlie the storm updraft and elevated charge dipole region. This proposition is exemplified by the Kenton downburst case discussed in this paper in which peak downburst winds, the bow echo apex, and maximum echo top heights were concurrently located at the observing station. It is apparent that peak downburst winds observed at the surface are believed to be associated with an elevated electric charge dipole resulting from the intense storm updraft, thus providing a tentative explanation of the relationship between CG lightning discharge and downburst observation at the surface. Most importantly, the analysis presents a readily identifiable signature: microscale regions of suppressed CG lightning activity (termed lightning "gap") associated with downburst occurrence. Validation efforts during future convective seasons over the High Plains should serve to further establish a relationship between thermodynamic environments, downburst occurrence and electrification of convective storms.

## 5. References

Atkins, N.T., and R.M. Wakimoto, 1991: Wet microburst activity over the southeastern United States: Implications for forecasting. *Wea. Forecasting*, **6**, 470-482.

Brock, F. V., K. C. Crawford, R. L. Elliott, G. W. Cuperus, S. J. Stadler, H. L. Johnson and M. D. Eilts, 1995: The Oklahoma Mesonet: A technical overview. *Journal of Atmospheric and Oceanic Technology*, **12**, 5-19.

Doswell, C.A., 2001: Severe convective storms- An overview. *Severe Convective Storms*, AMS Meteor. Monogr. Series, **27**, 570 pp.

Fujita, T.T., 1971: Proposed characterization of tornadoes and hurricanes by area and intensity. SMRP Research Paper 91, University of Chicago, 42 pp.

Fujita, T.T., 1985: The downburst: microburst and macroburst. SMRP Research Paper 210, University of Chicago, 122 pp.

MacGorman, D.R., and W.D. Rust, 1998: The electrical nature of storms. Oxford Univ. Press, New York, 422 pp.

Oklahoma Climatological Survey, cited 2006: *Climate of Oklahoma*. [Available online at http://climate.ocs.ou.edu/.]

Pryor, K.L., and G.P. Ellrod, 2004: WMSI - A new index for forecasting wet microburst severity. *National Weather Association Electronic Journal of Operational Meteorology*, 2004-EJ3.

Pryor, K.L., and G.P. Ellrod, 2005: GOES WMSI - progress and developments. Preprints, *21st Conf. on Wea. Analysis and Forecasting*, Washington, DC, Amer. Meteor. Soc.

Pryor, K.L., 2006: The GOES Hybrid Microburst Index. Preprints, *14th Conf. on Satellite Meteorology and Oceanography*, Atlanta, GA, Amer. Meteor. Soc.

Przybylinski, R.W., 1995: The bow echo. Observations, numerical simulations, and severe weather detection methods. *Wea. Forecasting*, **10**, 203-218.

Rakov, V.A., and M.A. Uman, 2003: Lightning: physics and effects. Cambridge University Press, New York, NY, 687 pp.

Saunders, C.P.R., 1993: A review of thunderstorm electrification processes. *J. Appl. Meteor.*, **32**, 642-655.

Stull, R.B., 1988: An introduction to boundary layer meteorology. Kluwer Academic Publishers, Boston, 649 pp.

Uman, M.A., 2001: The lightning discharge. Dover Publications, Mineola, New York, 377 pp.

Wakimoto, R.M., 1985: Forecasting dry microburst activity over the high plains. *Mon. Wea. Rev.*, **113**, 1131-1143.

Wakimoto, R.M., 2001: Convectively Driven High Wind Events. *Severe Convective Storms*, AMS Meteor. Monogr. Series, **27**, 570 pp.

**Acknowledgements**

The author thanks Jaime Daniels (NESDIS) and Raytheon contractors for providing GOES sounding retrievals displayed in this paper and Shuang Qiu (NESDIS) for providing NLDN processing and display software. The author also thanks Mr. Derek Arndt (Oklahoma Climatological Survey) and the Oklahoma Mesonet for the surface weather observation data used in this research effort. Cloud-to-ground lightning data was available for this project through the courtesy of Vaisala, Inc.